\documentclass[aps,twocolumn,amsfonts,amsmath,amssymb,showkeys,nobibnotes,nofootinbib,floatfix,a4paper,notitlepage]{revtex4-1}
\usepackage{graphicx} 
\usepackage{cleveref} 
\usepackage{color}
\numberwithin{equation}{subsection}

\makeatletter
\let\cat@comma@active\@empty
\makeatother

\begin{document}

	\title{When does a disaster become a systemic event?\\ Estimating indirect economic losses from natural disasters} 
	
	\author{Sebastian Poledna$^{1,2,3}$}
	\author{Stefan Hochrainer-Stigler$^{1}$}
	\author{Michael Gregor Miess$^{1,2,5,7}$}
	\author{Peter Klimek$^{2,3}$}
	\author{Stefan Schmelzer$^{5,7}$}
	\author{Johannes Sorger$^{2}$}
	\author{Elena Shchekinova$^{1}$}
	\author{Elena Rovenskaya$^{1,2,6}$}
	\author{Joanne Linnerooth-Bayer$^{1}$}
	\author{Ulf Dieckmann$^{1,2}$}
	\author{Stefan Thurner$^{1,2,3,4}$}

	\affiliation{$^1$International Institute for Applied Systems Analysis, Schlossplatz 1, 2361 Laxenburg, Austria} 
	\affiliation{$^2$Complexity Science Hub Vienna, Josefst{\"a}dter Stra{\ss}e 39, 1080 Vienna, Austria}
	\affiliation{$^3$Section for Science of Complex Systems, Medical University of Vienna, Spitalgasse 23, 1090 Vienna, Austria}
	\affiliation{$^4$Santa Fe Institute, 1399 Hyde Park Road, Santa Fe, NM 87501, USA} 
	\affiliation{$^5$Institute for Advanced Studies, Josefst\"adter Stra{\ss}e 39, 1080 Vienna, Austria}
	\affiliation{$^6$Faculty of Computational Mathematics and Cybernetics, Lomonosov Moscow State University (MSU), Moscow, Russia}
	\affiliation{$^7$Vienna University of Economics and Business, Welthandelsplatz 1, 1020 Vienna, Austria}

	\begin{abstract}
	Reliable estimates of indirect economic losses arising from natural disasters are currently out of scientific reach. To address this problem, we propose a novel approach that combines a probabilistic physical damage catastrophe model with a new generation of macroeconomic agent-based models (ABMs). The ABM moves beyond the state of the art by exploiting large data sets from detailed national accounts, census data, and business information, etc., to simulate interactions of millions of agents representing \emph{each} natural person or legal entity in a national economy. The catastrophe model introduces a copula approach to assess flood losses, considering spatial dependencies of the flood hazard. These loss estimates are used in a damage scenario generator that provides input for the ABM, which then estimates indirect economic losses due to the event. For the first time, we are able to link environmental and economic processes in a computer simulation at this level of detail. We show that moderate disasters induce comparably small but positive short- to medium-term, and negative long-term economic impacts. Large-scale events, however, trigger a pronounced negative economic response immediately after the event and in the long term, while exhibiting a temporary short- to medium-term economic boost. We identify winners and losers in different economic sectors, including the fiscal consequences for the government. We quantify the critical disaster size beyond which the resilience of an economy to rebuild reaches its limits. Our results might be relevant for the management of the consequences of systemic events due to climate change and other disasters. 
	\end{abstract}


\maketitle 

\section{Introduction}
Total economic losses from natural and man-made disasters in 2017 are estimated to be USD 306 billion.\footnote{\url{http://www.swissre.com/media/news_releases/nr20171220_sigma_estimates.html}} With hurricanes Harvey, Irma and Maria, which have made 2017 the second costliest hurricane season on record, the US was hit particularly hard. Due to climate change, economic losses from extreme events -- such as floods, droughts, and other climatic disasters -- will further increase, and events that have been considered rare until now will become more common in the future \citep{Melillo:2014aa}.
However, the resulting economic losses to a national economy are difficult to quantify.
While direct losses due to the immediate destruction of homes, firms, infrastructures and lives can be reliably estimated with existing data, estimates of indirect losses -- which arise as economic consequences of physical destruction -- are much harder to obtain, and no consensus on their validity exists \citep{National-Research-Council:1999aa}. The primary reason for this is that natural disasters affect the economy in multiple ways and along several dimensions \citep{Hallegatte:2010aa,linkov2014changing}. For example, a flood event might destroy or damage the physical capital of a firm, causing output losses and worker layoffs. This, in turn, can trigger further output losses to suppliers and customers of that firm, potentially leading to more layoffs. The associated reduction in household income additionally lowers consumption, potentially further enhancing output losses and increasing layoffs. Subsequently, however, the reconstruction of damaged or destroyed capital can have the opposite effect: companies involved in reconstruction activities experience growth and expand their workforce, which results in higher income that ripples through the economy and leads to economic growth via Keynesian multiplier effects. 
Hence, natural disasters simultaneously cause both indirect losses and indirect gains. We refer to these losses and gains as indirect economic effects.\footnote{To avoid double counting, we measure losses or gains as the change in gross domestic product (GDP) relative to a baseline scenario. This definition differs from \cite{Cochrane:2004aa}, who additionally attributes employment losses (e.g. due to the closure of damaged facilities) to direct losses, and defines indirect losses as all economic consequences except for damages (direct losses) and employment losses caused by the disaster.} 

Indirect economic effects of average natural disasters, as measured by changes in GDP, are typically small. Net effects may even be close to zero in the short term, where losses and gains of natural disasters can cancel each other out \citep{West:1996aa}. Still, there are winners and losers, since effects may differ substantially across industries and economic sectors, and almost never cancel for companies and individuals \citep{Loayza:2012aa}. Several empirical studies even find small but positive short- to medium-term overall effects for moderate natural disasters of certain types, especially for moderate floods \citep{Fomby:2013aa,Loayza:2012aa,cavalloetal2013,cunado2014macroeconomic,cavallo2011natural,kousky2014informing,shabnam2014natural,leiter2009creative}. However, the evidence on the sign and magnitude of indirect economic effects of natural disasters in general remains mixed and conflicting, see \cite{cavallo2011natural,kousky2014informing,shabnam2014natural} for recent surveys. This situation is echoed in meta-analyses such as \cite{lazzaroni2014natural}, which reports that disasters on average have an insignificant impact in terms of indirect costs, and \cite{klomp2014natural}, which finds some evidence that a part of the negative impact of natural disasters reported in studies is caused by a publication bias. Moreover, \citep{mohan2018decomposing} shows that the impacts of natural disasters on different components of GDP (such as investment, government and private consumption, exports, imports) differ widely in timing, direction and extent. More aggregate analyses might mask these differences, and it may be difficult to find clear and large net aggregate impacts on GDP.
Severe natural disasters (systemic events), on the other hand, are not believed to have neutral or positive economic effects on an aggregate level \citep{Fomby:2013aa,cavalloetal2013,Loayza:2012aa,kousky2014informing,noy2009macroeconomic,hochrainer2009}. Indirect economic losses from these systemic events may be amplified by several mechanisms such as post-disaster inflation, network effects like supply chains affecting firms that were not initially impacted by the event, as well as physical and financial resource constraints regarding the productive capacity of the economy \citep{Hallegatte:2010aa}. Despite these general arguments, empirical evidence on indirect economic effects of systemic events, particularly regarding severe floods in developed economies, is scarce and largely inconclusive \citep{Fomby:2013aa,Loayza:2012aa,cunado2014macroeconomic,noynualsri2007,jaramillo2009natural,cavalloetal2013}.

Indirect economic effects from natural disasters are difficult to model with traditional approaches such as input-output (IO) models \cite{okuyama2014disaster,hallegatte2008adaptive,okuyama2007economic,Rose:1997aa,Whitman:1997aa,Boisvert:1992aa,Cochrane:1997aa}, computable general equilibrium (CGE) models \cite{Strulik:2016aa,narayan2003macroeconomic,RISA:RISA912,rose2005modeling,Brookshire:1992aa,Boisvert:1995aa}, and econometric analyses \citep{Fomby:2013aa,cavalloetal2013,Loayza:2012aa,cunado2014macroeconomic,kousky2014informing,noy2009macroeconomic,hochrainer2009,cavallo2011natural,leiter2009creative,raddatz2009,kousky2014informing,shabnam2014natural,mohan2018decomposing,lazzaroni2014natural,klomp2014natural} because of the over-simplifying nature of these approaches \citep{rose2004}. 
IO models have a tendency to overestimate indirect economic losses and gains \citep{National-Research-Council:1999aa}. This was demonstrated, for example, 
for the case of job gains from reconstruction following the Northridge earthquake in California in 1994: \cite{Bolton:1995aa} showed that the IO estimates by \citep{cochrane1996} significantly exceeded actual data for the Los Angeles area following the earthquake \citep{National-Research-Council:1999aa}.
Furthermore, IO models cannot incorporate the reactions of economic agents to a disaster. By design, CGE models are overly optimistic regarding the flexibility of an economy to react to natural disasters \citep{Hallegatte:2010aa}. 
This is due to the underlying assumption that price clearing mechanisms bring the economy back to a general equilibrium after a certain time and are subject only to modest constraints such as adjustment costs in the investment function. However, real-world price formation is sticky and imperfect. 
Production functions typically assumed in CGE models\footnote{Constant elasticity of substitution (CES) functions.} often overstate the flexibility of substitution between factors of production (labor, capital, material input). Econometric analysis of indirect economic effects faces the problem that statistical, historical relationships used to derive model parameters are likely to be disrupted by the disaster \citep{National-Research-Council:1999aa}.
The latter study points out an array of factors that may challenge the implicit socio-economic assumptions of the approach, including: the temporary nature of measures taken directly after the disaster event, permanent economic changes (e.g. in the production function), purchase and sale patterns, as well as labor force migration or overtime hours in the reconstruction phase. Indicative of this, \cite{West:1996aa} demonstrates that the regional econometric model results in \cite{West:1994aa} over-estimated the economic impact of Hurricane Andrew, which hit the US in 1992, by 70-85 \% \citep{National-Research-Council:1999aa}.

In summary, traditional approaches so far have failed to provide unambiguous conclusions about the indirect economic effects of natural disasters, in particular regarding systemic events.
This means that, given the characteristics of these traditional approaches, today it is still largely impossible to relate the size of initial damages of natural disasters to the expected subsequent indirect economic effects, or to clearly disentangle expected losses and growth effects on a sectoral level. To address this problem, we propose a novel approach that combines a probabilistic physical damage catastrophe model with a new generation of macroeconomic agent-based models (ABMs). The ABM we adopt moves beyond the state of the art by exploiting large data sets from detailed national accounts, census data, and business information, etc., to simulate the interactions of millions of agents representing each natural person or legal entity, such as corporations, government entities and institutions in a national economy. It has been shown that this model is able to forecast numerous macroeconomic variables including major variables such as GDP, inflation, consumption and investment better than standard forecasting approaches \citep{Poledna:2017ab}.

\begin{figure}
	\begin{center}
		\includegraphics[width=.49\textwidth]{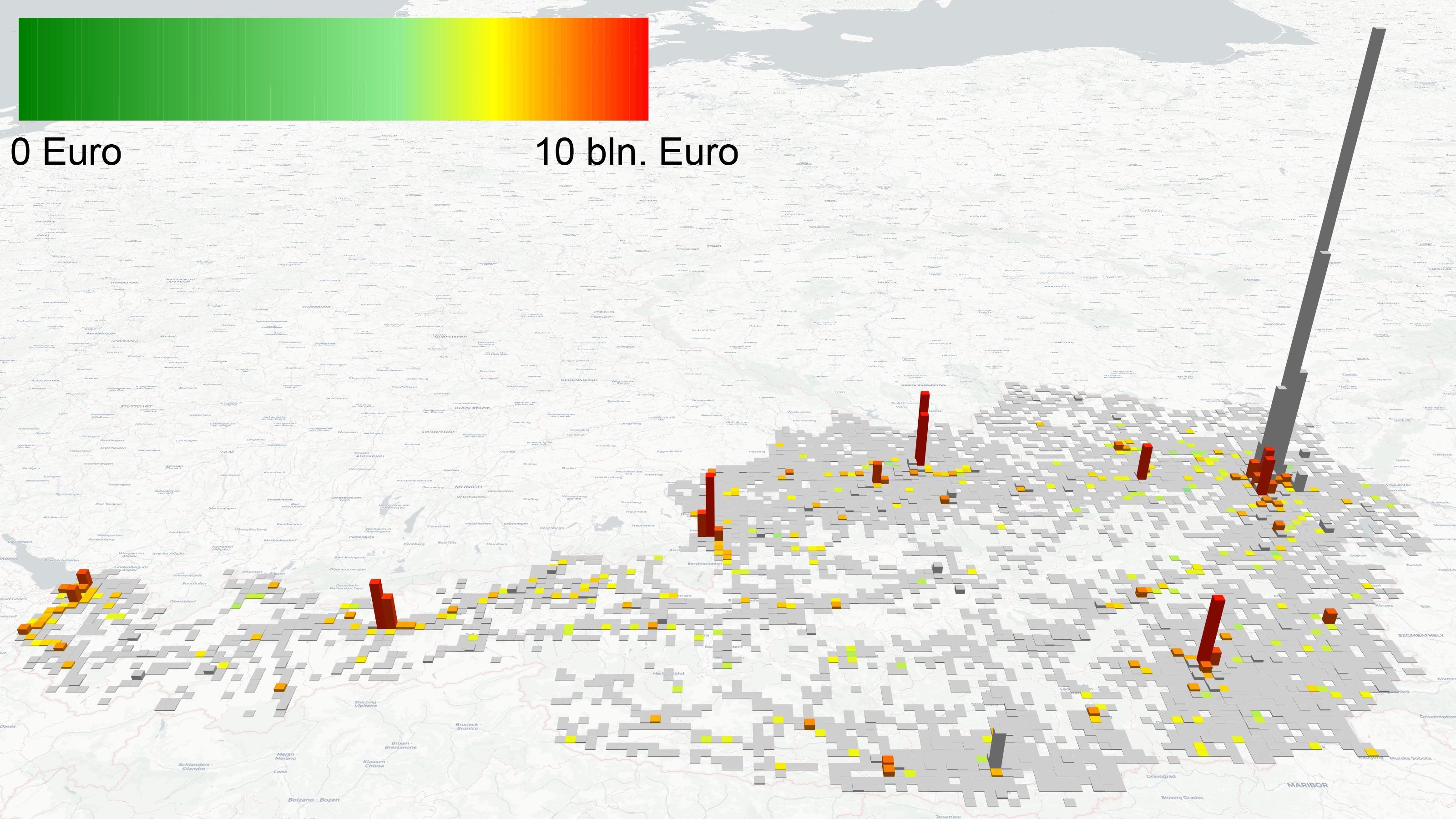}
	\end{center}
\caption{Geospatial distribution of capital across Austria. Affected capital in flooding zones of a 250-year event is shown in a color scale from green to red, where green indicates small damages and red large damages up to 10 bln. Euro. Gray indicates capital stock not affected by the flood. For example, capital stock in Vienna (the large bar in the northeast) is not affected due to extensive flood protection measures, and is therefore shown as grayed out.}
\label{fig:generator}
\end{figure}

We apply our method to estimate indirect economic losses from flood events in Austria, for which the mentioned data is available in highly detailed, complete and consistent form (see SI\footnote{\url{http://www.complex-systems.meduniwien.ac.at/people/spoledna/supporting_information_poledna_et_al_2018.pdf}}). The ABM is geolocalized in the sense that the location of capital stock, e.g. firms and infrastructure, is known. The geospatial distribution of capital across Austria is shown in Fig. \ref{fig:generator}. The physical damage catastrophe model (damage scenario generator) allows us to generate realistic virtual natural disaster events (floods) of controlled size. It is based on a copula approach that assesses realistic hazard frequency and intensity, and takes into account spatial dependencies and the dependency on the severity of the events (see SI). After generating a geolocalized flood event of a given size at the beginning of year 2014,\footnote{This year is chosen to simulate the flooding event since 2013 it is the last year for which the main data source -- the symmetric IO tables for the Austrian economy -- is available to calibrate the model.} the affected dwellings, firms and infrastructure are destroyed (the affected capital in flooding zones of a 250-year event is depicted in Fig. \ref{fig:generator}). On this basis, the detailed indirect economic effects across economic sectors and industries are studied in the ABM over several consecutive years.

\section*{Results}
\subsection*{Moderate disasters do not always have a negative impact on economic growth}
\begin{figure}
	\begin{center}
		\includegraphics[width=.49\textwidth]{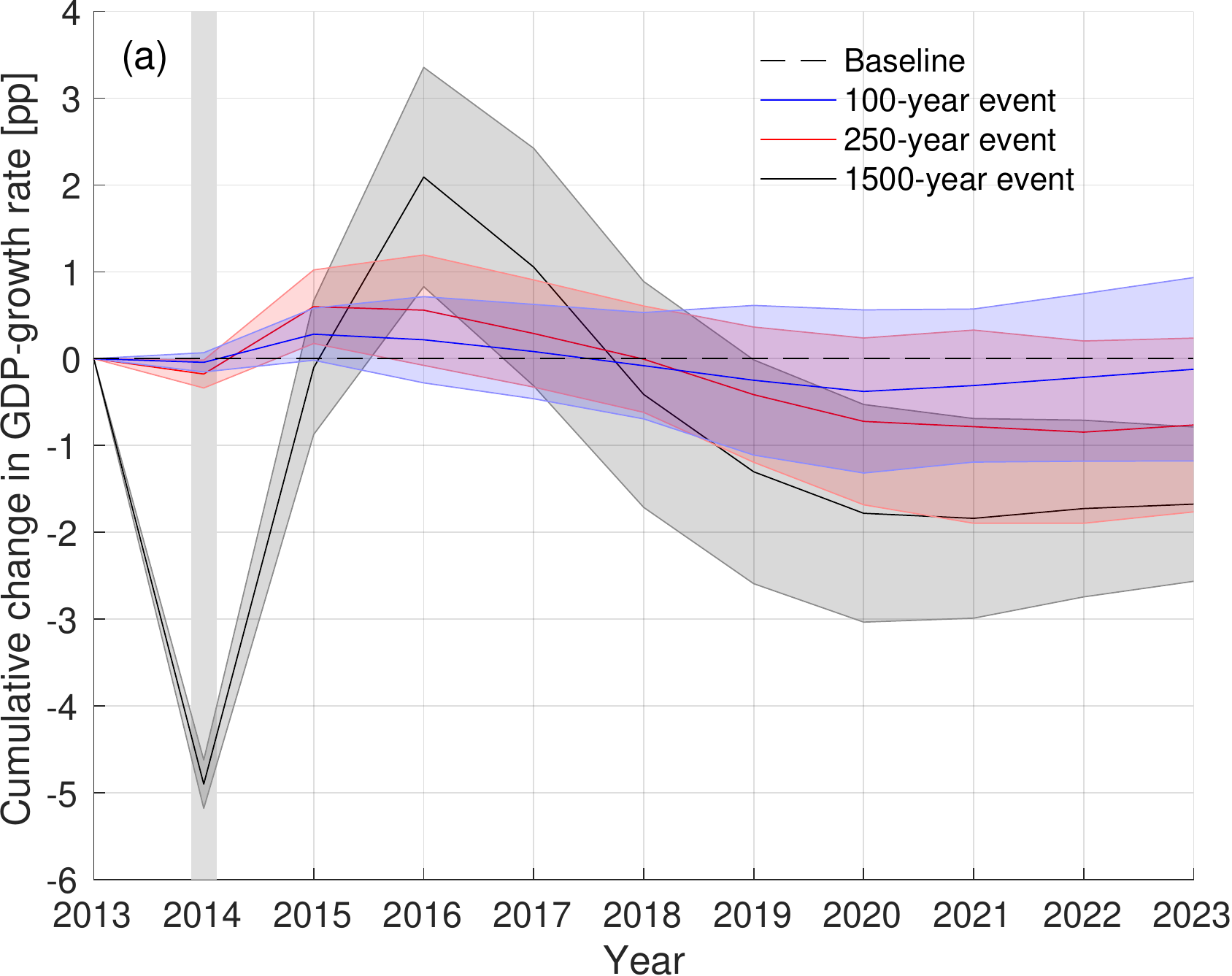}
		\includegraphics[width=.49\textwidth]{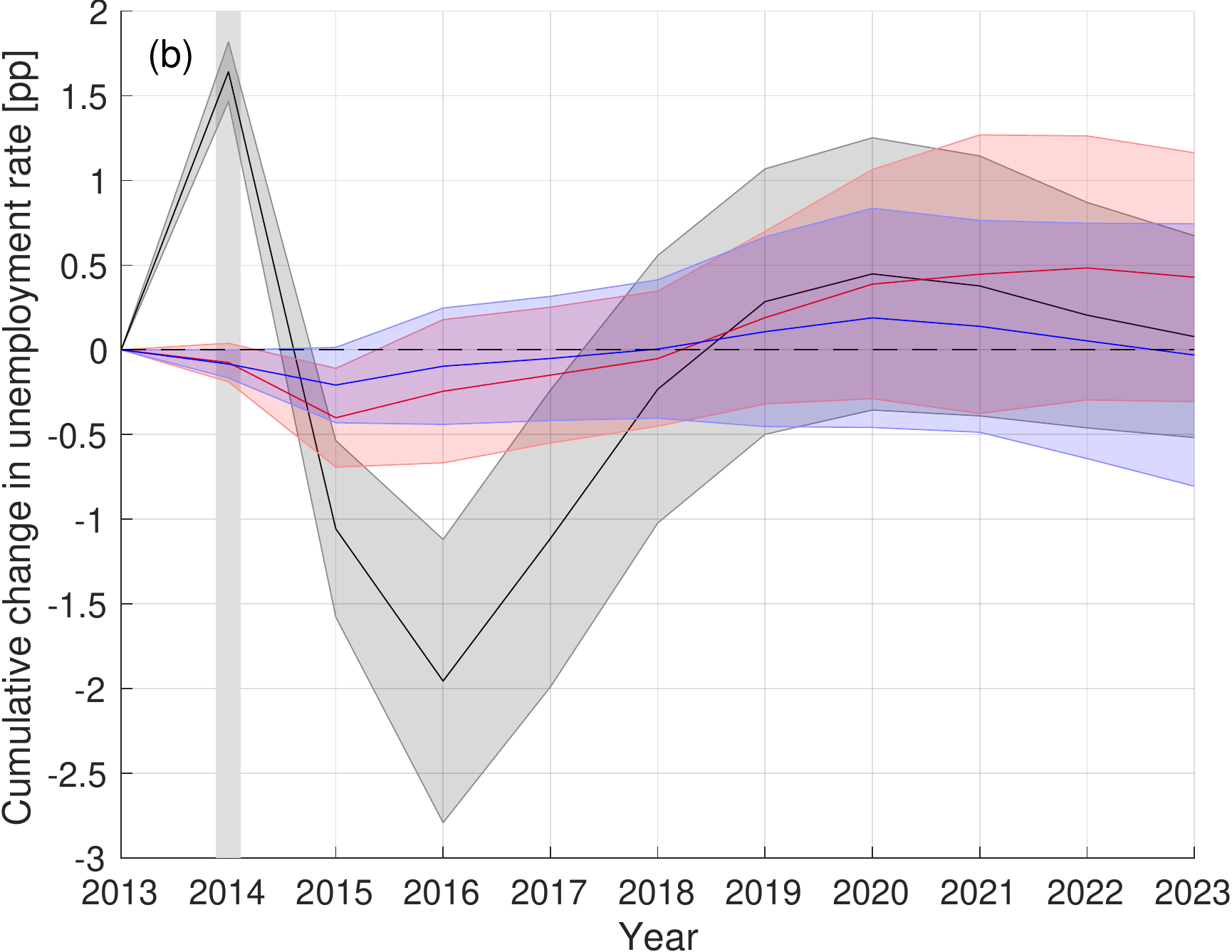}
		\includegraphics[width=.49\textwidth]{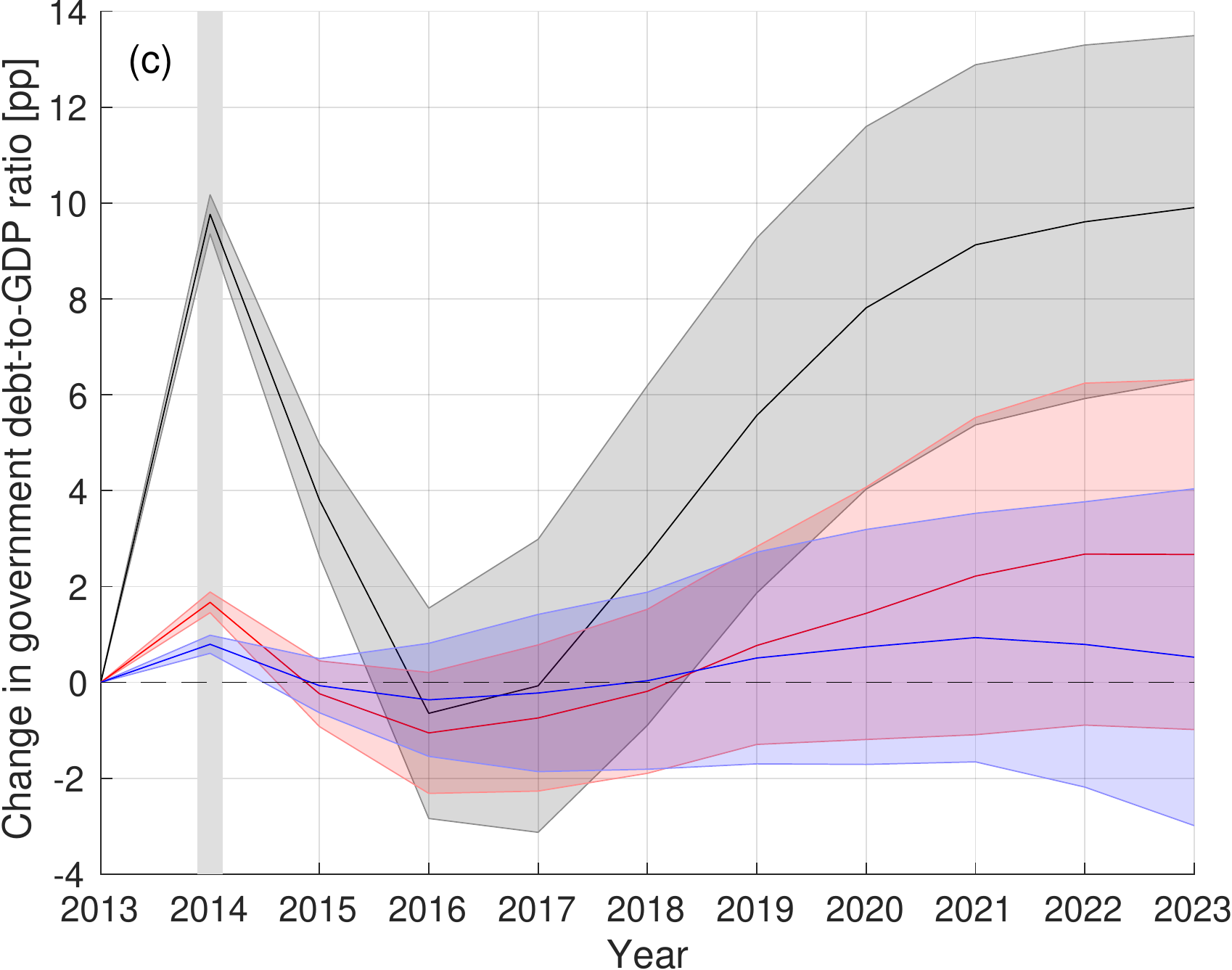}
	\end{center}
\caption{Indirect economic gains and losses of a 100- (\textcolor{blue}{blue}), 250- (\textcolor{red}{red}) and 1,500- (\textcolor{black}{black}) year flood event. Time labels on the x-axis indicate the end of each year, and the gray vertical bar marks the first year after the flood. The panels show the effects as percentage point changes relative to the baseline scenario, in which no disaster happens.  
(a) Cumulative changes in GDP-growth rates. 
(b) Cumulative changes in the unemployment rate. 
(c) Changes in the government debt-to-GDP ratio. 
Shaded areas cover one standard deviation above and below the mean values, as obtained from 50 independent Monte-Carlo simulations.}
\label{fig:macro_effects}
\end{figure}
Fig. \ref{fig:macro_effects} shows the indirect economic effects resulting from a 100-year (blue line) and a 250-year (red line) flood event that destroys dwellings and productive capital.\footnote{The event occurs at the beginning of the year 2014.} The total direct losses (damages) amount to about 0.6 \% (100-year event) and 1.2 \% (250-year event) of Austrian capital stock, respectively. Fig. \ref{fig:macro_effects}(a) shows the cumulative change in the GDP growth rate relative to the baseline scenario\footnote{The baseline scenario describes a continuation of current trends for the Austrian economy. The baseline scenario serves as the benchmark against which we evaluate the indirect economic effects of the different flooding scenarios.} in percentage points (pp).\footnote{A percentage point (pp) is the unit for the arithmetic difference of two percentages. For example, moving up from 10\% to 12\% is a 2 pp increase, but it is a 20 percent increase in what is being measured.} The qualitative behavior of the two scenarios is similar: starting from small negative effects immediately after the disaster, cumulative effects on economic growth then become positive in the short to medium term (2014-2018), and turn negative in the long term. These effects are most pronounced with an about 0.5 pp cumulative GDP growth rate increase (250-year event) relative to the baseline scenario in the second year after the flood (2015). In the long term, primarily due to a multiplier-accelerator mechanism \cite{samuelson1939} (see below), the effects decline to an almost neutral impact (100-year event), and to a reduction of GDP growth of approx. 0.8 pp (250-year event), respectively. This behavior, i.e. positive short- to medium-term and negative long-term effects of moderate size, is in line with the literature \citep{Fomby:2013aa,Loayza:2012aa,cunado2014macroeconomic,leiter2009creative,raddatz2009}. Fig. \ref{fig:macro_effects}(b) demonstrates that -- as to be expected according to Okun's law -- the change in the unemployment rate is inversely correlated to economic growth, but at a slightly lower amplitude: for the 250-year event, a cumulative decline of almost 0.5 pp two years after the flood (2015) is followed by a cumulative growth in the unemployment rate up to a maximum of about 0.5 pp in the long term. Fig. \ref{fig:macro_effects}(c) depicts the government debt-to-GDP ratio and shows that the dynamics of the growth and unemployment rates, as well as the transfer we assume to be provided by the government to fully compensate households for their losses of dwellings as catastrophe relief, all lead to an initial rise in this ratio of about 2 pp. For three years after the flood (2015-2018), the government debt-to-GDP ratio temporarily falls slightly below its initial level, but in the long run stabilizes at an increase by more than 2 pp relative to the baseline scenario (250-year event).

\subsection*{Severe disasters have pronouncedly negative economic effects immediately after the event and in the long term} 
A severe-disaster scenario that simulates a 1500-year flood event is shown in Fig. \ref{fig:macro_effects} (black lines). The total direct losses correspond to approximately 10 \% of the capital stock in Austria. The indirect economic effects after this shock are qualitatively different from the moderate-disaster scenarios. The initial overall effect on GDP growth is pronouncedly negative, with a cumulative reduction of GDP growth by about 5 pp, see Fig. \ref{fig:macro_effects}(a). Due to reconstruction, growth picks up fast in the year after the disaster, and surpasses cumulative GDP growth of the baseline scenario by the second year after the flood (2015), culminating in a temporary economic boost of about 2 pp of additional cumulative GDP growth in 2016. The multiplier-accelerator mechanism \cite{samuelson1939}, as well as production, capacity and credit constraints (see SI) drag growth downwards after this point, leading to negative long-term cumulative growth effects of approx. 1.7 pp. 
The unemployment rate reacts strongly to the severe disaster, with a cumulative initial increase of more than 1.5 pp, and is followed by a reduction up to almost 2 pp in 2015 during the reconstruction phase, see Fig. \ref{fig:macro_effects}(b). After this pronounced disruption of the labor market, the unemployment rate rises again by a cumulative change of approx. 0.5 pp (2020) due to the cyclical dynamics, and stabilizes at a level close to the baseline scenario in the long term.
Immediately after the disaster, a large initial government transfer to households to compensate for their losses of housing stock,\footnote{We assume -- in line with past experiences of political processes regarding catastrophe relief by the Austrian government -- this transfer to be limited to about a third of the total losses in dwelling stock.} as well as substantial decreases in government revenues and GDP, lead to a 10 pp rise of the government debt-to-GDP ratio, see Fig. \ref{fig:macro_effects}(c). Even though this ratio shortly returns to its initial level due to the positive economic effects of reconstruction, the downturn because of over-production three years after the flood event implies a subsequent rise in this ratio by almost 10 pp, leaving government finances substantially deteriorated in the long term. 

\noindent
\subsection*{Effects differ substantially across industries and economic sectors}
\begin{figure}
	\begin{center}
		\includegraphics[width=.49\textwidth]{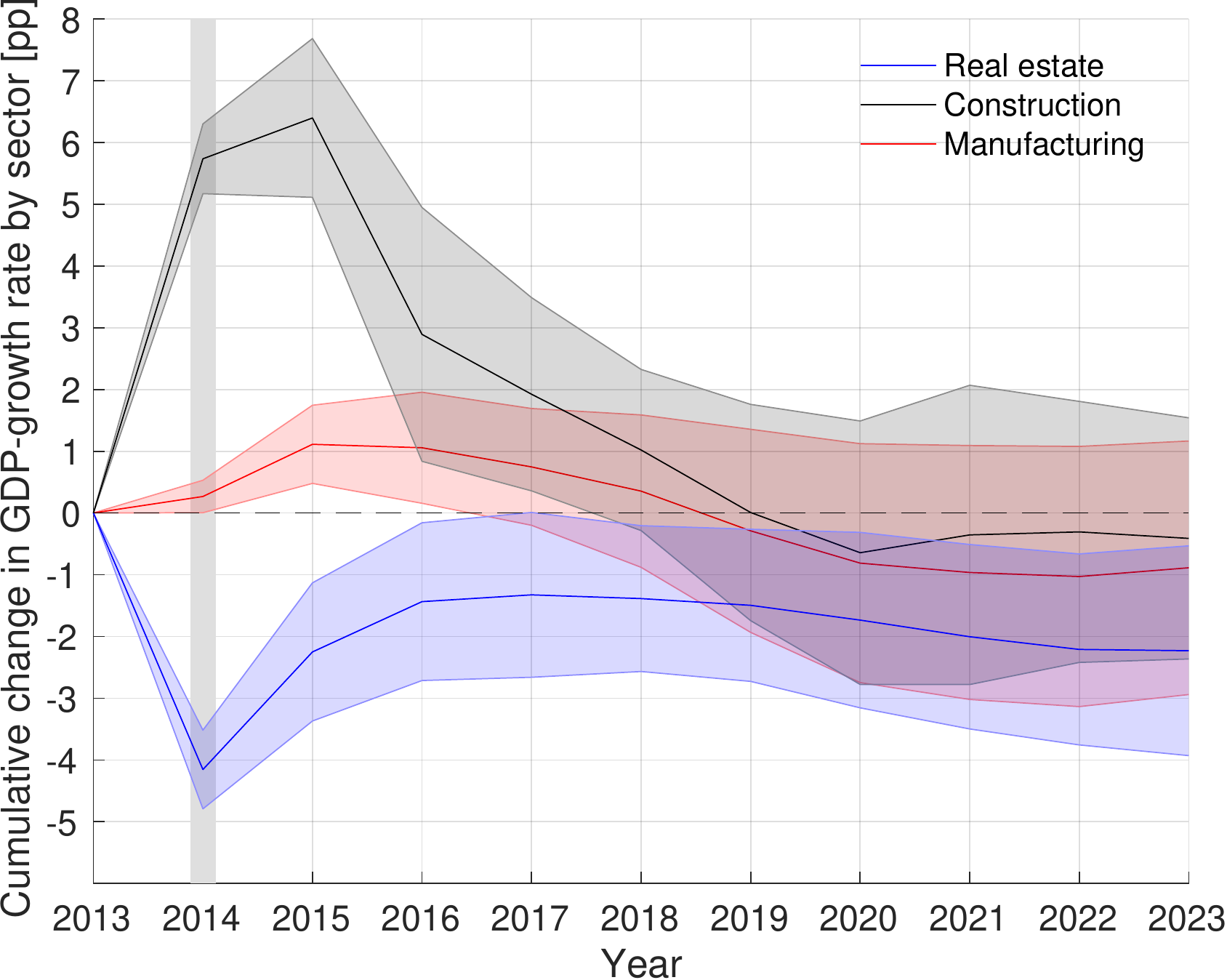}
	\end{center}
\caption{Cumulative growth effects on sectoral GDP after a 250-year flood event for selected economic sectors in percentage points (pp) relative to the baseline scenario. Shaded areas cover one standard deviation above and below the mean values. Sectors shown: construction sector (black), manufacturing sector (\textcolor{red}{red}) and real estate sector (\textcolor{blue}{blue}). The gray vertical bar indicates the year of the flood.}
	\label{fig:cumgva250}
\end{figure}

While moderate flood events can have positive aggregate effects in the medium term, impacts are expected to differ significantly across economic sectors. Fig. \ref{fig:cumgva250} confirms this conjecture. It shows the effects for the most severely impacted sectors as a result of the 250-year event. The real estate sector (blue line) suffers substantially from the destruction of residential capital stock: sectoral output is reduced by more than 4 pp. Despite reconstruction works improving the initial situation, the cumulative growth change in this sector remains negative, with a cumulative loss in growth of about 2 pp in the long run. The construction sector (black line) immediately profits from the reconstruction of dwellings and productive capital with a sectoral GDP growth of almost 6 pp in the first year after the flood (2014). After the fast ramp-up of reconstruction during the first years after the flood, peaking in an about 6.5 pp increase in the second year after the flood (2015), this effect gradually wears off in the following years, turns slightly negative by year seven after the flood (2020) and remains rather stable at this level thereafter. The restoration of productive capital takes more time. The largest cumulative increase for the manufacturing industry (red line) of about 1 pp is reached in year two after the flood (2015), since this sector supplies a major part of the material input for the re-installment of losses in productive capital. Following the general downturn due to over-production and the thereby induced economic cycle, we see that cumulative GDP growth of this sector in the long term (2019-2023) is lower by almost 1 pp than in the baseline scenario. The effects on all sectors can be found in Table S7 in the SI.

\subsection*{Loss of resilience -- when disasters become systemic events}
\begin{figure}
	\begin{center}
		\includegraphics[width=0.49\textwidth]{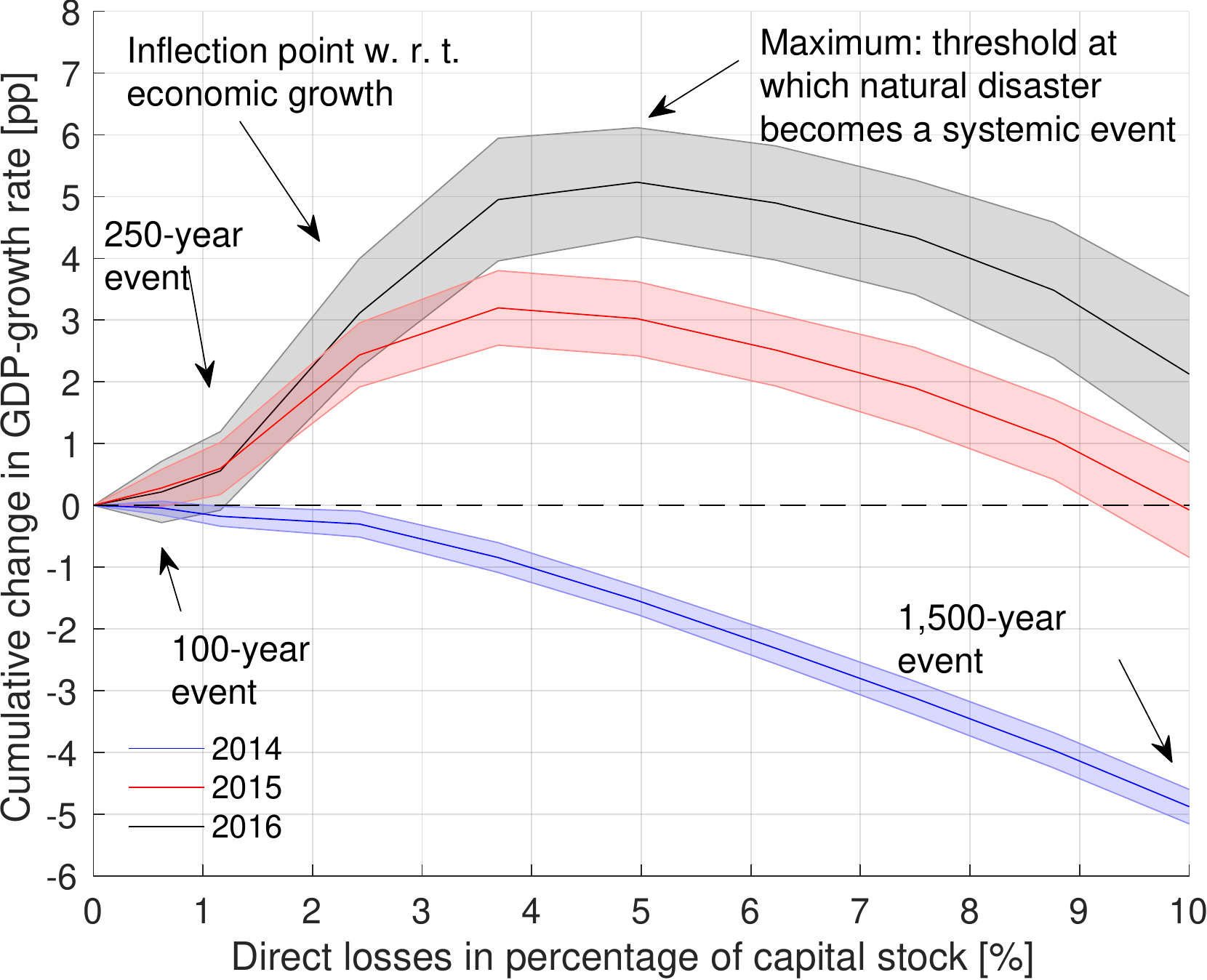}
	\end{center}
	\caption{Cumulative changes in GDP growth relative to the baseline scenario as a function of the direct damage as a percentage of GDP. Results are shown for three different years after the disaster: \textcolor{blue}{2014}, \textcolor{red}{2015} and 2016. Shaded areas cover one standard deviation above and below the mean values. Immediately after the event (2014), all disaster sizes are associated with negative growth relative to the baseline scenario. In contrast, for the years 2015 and 2016 there exist inflection points and maxima for GDP growth, indicating the existence of direct damage sizes that, respectively, are ``optimal'' in terms of economic growth and determine a threshold where natural disasters become systemic events.}
	\label{fig:stefansfigure}
\end{figure}
Fig. \ref{fig:stefansfigure} shows the cumulative changes in GDP growth (relative to the baseline scenario) as a function of the sizes of direct damages of the event for different times after the disaster: one year (2014), two years (2015), and three years (2016) after the flood event. 
As can be expected, the larger the direct damage, the larger the decline of GDP growth immediately after the flooding disaster (2014) is. In the second year after the flood event (2015), the effects of increased economic activity induce positive overall GDP growth. Within this second year, a large part of the initial damage can already by compensated, i.e. cumulative growth effects in the second year are slightly positive, but remain below 1 \% for disasters smaller than a 250-year event. For larger events, the economy shows a remarkable growth that is induced by reconstruction after the disaster, which clearly outweighs the direct losses. This growth is limited by different constraining factors (see SI) and starts to decline with respect to the direct losses inflicted by the disaster: the inflection point of the curve occurs at about a 2.4 \% loss of capital stock. The cumulative changes in GDP growth in this year show a maximum in the region where the initial damage causes about 3-4 \% direct losses in the capital stock. 
At this maximum, the growth effects lose momentum and indirect losses start to dominate beyond the maximum. The situation is similar for two years after the event (black line): while the position of the inflection point remains at about 2.4 \%, the growth stimulus is somewhat more pronounced, and the maximum 
is located at approximately 5 \% direct losses. At this point economic growth is severely restrained, the ability for resilience is lost, and the natural disaster becomes a systemic event.

\section*{Discussion}
\label{sec:discussion}
We present a novel approach for estimating the indirect economic losses caused by natural disasters by combining a probabilistic physical damage catastrophe model with a macroeconomic ABM. 
The method has been applied to flood scenarios in Austria. The ABM is calibrated to the Austrian economy (households, non-financial and financial firms and a government sector) at a scale of 1:1, i.e. every economic agent in Austria (about 10 million) is represented in the model. The ABM incorporates an input-output model with 64 industries, where all goods and services are produced endogenously, and depicts the relevant constraints regarding productive capacities and financing conditions at the level of individual agents. This allows us to estimate the indirect economic effects of natural disasters that are caused by the unfolding sequence of economic events following the initial destruction of productive capital and dwellings. In particular, the model shows the indirect economic effects of simulated disaster shocks of controlled size on the Austrian economy in terms of cumulative GDP growth and other major macroeconomic variables

The model produces realistic results in a number of aspects. Specifically, results correspond well with recent empirical findings of studies including sectoral detail and different types of disasters such as \cite{Fomby:2013aa,Loayza:2012aa,cunado2014macroeconomic,leiter2009creative,noy2009macroeconomic,raddatz2009}. We find that moderate losses due to 100- and 250-year flood events have small but positive short- to medium-term impacts, while they lead to negative long-term effects of similar magnitude.
These results correspond particularly well with \cite{cunado2014macroeconomic}, which shows that floods of moderate magnitude in developed economies, while resulting in positive short- to medium-term effects, have slightly negative cumulative effects in the long run. Short- to medium-term results are further supported by a firm-level empirical study on flooding disasters in Europe \citep{leiter2009creative}, which finds higher average firm asset and employment growth for regions affected by flooding disasters. Comparable impacts are also obtained by \cite{Loayza:2012aa,Fomby:2013aa}, which report aggregate short- to medium-term positive growth effects of floods, and by \cite{noy2009macroeconomic}, which shows that natural disasters have positive short-term growth effects in developed economies. Negative long-term effects for climatic disasters are also reported in \cite{raddatz2009}. 

Our study is the first to estimate the indirect economic impacts of severe floods in a developed economy on an empirical basis, at the same time taking account of complex economic interactions and dynamics with our modeling approach. Simulations of severe disasters, such as a 1,500-year flood event in Austria, 
induce a pronouncedly more negative long-term economic impact on the Austrian economy. These results, in line with a theoretical analysis conducted by \cite{hallegatte2007economic}, demonstrate that negative indirect effects from severe disasters are primarily due to constraints on the productive capacity of the existing capital stock, credit provision and government finances, which impede immediate reconstruction and thus impose a ceiling on positive growth effects.
Empirical findings regarding severe floods are largely inconclusive: \cite{cunado2014macroeconomic,Fomby:2013aa,Loayza:2012aa,noynualsri2007,jaramillo2009natural,cavalloetal2013} all report no, insignificant or (once controlled for) vanishing effects for systemic floods in particular. Especially regarding the most detailed study on flooding disasters \citep{cunado2014macroeconomic}, a lack of data on severe floods in developed countries seemed not to have permitted the obtaining of results for this country group.

A unique feature of this analysis is that disaster impacts are disaggregated across 64 industry sectors and simultaneously tracked over time, demonstrating how positive aggregate economic consequences may result in winners and losers subject to particular dynamics. The sectoral decomposition of results reveals that, while some sectors providing the means for the reconstruction of capital stock might profit (predominantly the construction sector, to a lesser extent the manufacturing sector), others that are particularly hit by the disaster suffer from large losses that take several years to be compensated (especially the real estate sector). We compute the distribution of losses and their dynamics over time across sectors at a higher level of detail than previous studies. Models featuring endogenous dynamics such as CGE models, even though they depict up to 35 industry sectors as in \cite{narayan2003macroeconomic}, typically are comparative static CGE models\footnote{I.e. they compare an initial equilibrium state before the disaster and another equilibrium state after the disaster without consideration of the dynamics between these two economic equilibria.} \cite{narayan2003macroeconomic,RISA:RISA912,rose2005modeling}, while fully dynamic CGE models usually depict only one output good as in \cite{Strulik:2016aa}. Furthermore, they often are confined to regions smaller than a national economy \citep{RISA:RISA912,rose2005modeling}. Sectoral empirical studies such as \cite{Fomby:2013aa,Loayza:2012aa,cunado2014macroeconomic} -- besides being limited methodologically as set forth above -- typically divide the economy into two to three aggregate sectors (agriculture, industry and/or service sector). IO models applied to comparable contexts, such as \cite{hallegatte2008adaptive}, usually exploit the full range of national IO tables (mostly around 60-70 industries), but lack the endogenous non-linear dynamics present in the ABM.

We show that disasters trigger cyclical economic responses that follow a classic multiplier-accelerator mechanism as described in \citep{Samuelson:1939aa}. 
The cycle is caused by an overshooting of investment during the reconstruction phase leading to an economic boom, which is followed by a downturn due to a lack of demand once the restoration of capital stock has been completed. As reported in \cite{Fomby:2013aa,Loayza:2012aa,cunado2014macroeconomic}, indirect economic effects immediately after the event are predominantly negative due to initial losses of capital stock, income, as well as demand and supply of goods. The positive economic stimulus due to reconstruction occurs with a delay of at least a quarter, because some time is necessary to compensate for lost capital stock and income. This impulse in turn triggers an economic cycle. In the long term, the consequences of this cycle tend to outweigh the positive economic impact induced by reconstruction activities. Such a cyclical mechanism -- which is different from Schumpeter's creative-destruction or productivity effect, which has been the focus of several studies\footnote{We do not consider this productivity effect in the present study, since its empirical relevance is unclear and subject to extensive debate in the literature, where different empirical studies present mixed evidence on growth effects and associated increases in capital productivity after a natural disaster. For further discussion on the productivity effect see \cite{Hallegatte:2010aa}, for empirical studies presenting positive growth effects attributed to the productivity effect see \cite{albalabertrand1993}  and \cite{skidmoretoya2002}, whose findings are contradicted by several other empirical studies, see \cite{noynualsri2007,noy2009,hochrainer2009,jaramillo2009,raddatz2009}.} -- has received little attention in the literature up to now, with the exception of a theoretical analysis in \cite{Hallegatte:2008aa}.

This study is the first to combine a probabilistic physical damage catastrophe model with a macroeconomic ABM to quantitatively relate disaster sizes with the indirect economic impacts. We find a non-trivial behavior of the cumulative GDP growth effects as a function of the direct damage size. We determine a threshold beyond which the full productive capacity of an economy has been exploited to restore destroyed capital stock. At this point, which is at around 5 \% of destroyed capital stock, resilience is lost, and economic growth is dominated by direct losses. Previous studies such as \cite{hallegatte2007economic} have hitherto investigated this matter on a theoretical and more aggregate basis only.

We believe that in times of increased frequency and severity of potential climate-change-related natural disasters, it is expedient to anticipate their short- to long-term economic implications, direct as well as indirect ones. 
In particular, it is important to identify potential economic losers of these events, so as to optimally prepare for a fair and efficient post-crisis management.   

\section*{Materials and methods}

\subsection*{Agent based model of a small economy}
\label{subsec:abm}
\begin{figure}
	\begin{center}
		\includegraphics[width=.49\textwidth]{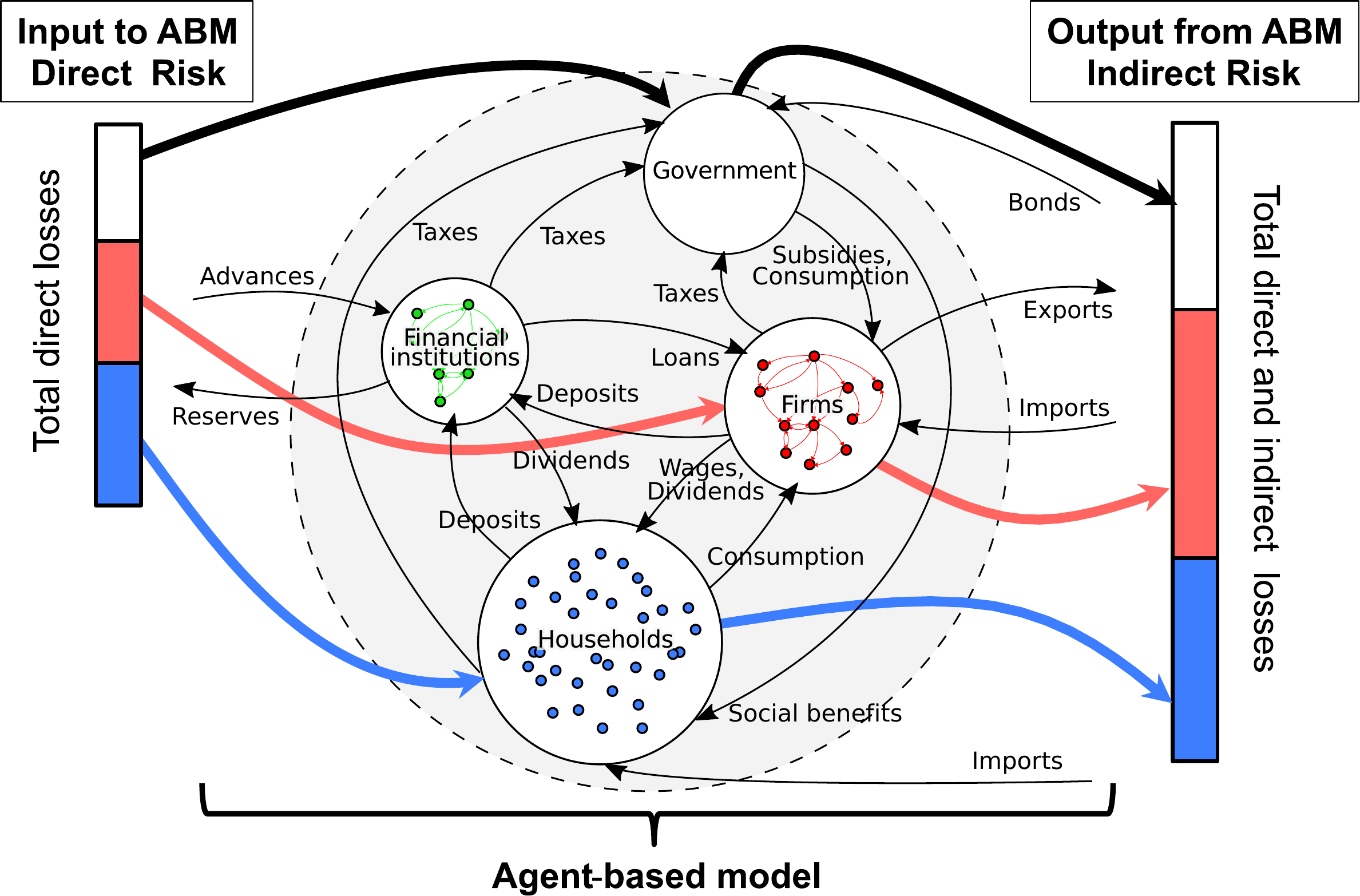}
	\end{center}
	\caption{Schematic overview of the ABM structure showing the institutional sectors (households, non-financial and financial firms and a general government), and their interactions. The stacked bars show an example of the distributions of direct (left) and indirect (right) total losses to the government (white), firms (red), and households (blue).} 
\label{fig:abm} 
\end{figure}
We employ an ABM -- developed in \cite{Poledna:2017ab} and \cite{Assenza:2015aa} -- which depicts the economy of a 
small nation at a 1:1 scale (about 10 million agents) (see SI for a model description).
The model is based on detailed data sources from national accounts, input-output tables, government statistics, census data and business surveys, and is able to closely approximate time series of major macroeconomic variables (GDP, inflation, household consumption, investment). The basic structure of the model is depicted in Fig. \ref{fig:abm}.
The model is calibrated to the economy of Austria in the year 2013, for which the required data is available (see SI). Large 
economies are still out of scope for such simulations within reasonable computing time. Simulations were carried out on the supercomputer of the Vienna Scientific Cluster. 

\subsection*{Flood risk estimation and damage scenario generator}
We estimate the disaster risk distributions for flood losses in Austria using a copula approach, and build a damage-scenario generator based on spatially explicit data to simulate losses to individual households, non-financial and financial firms and government entities across the 64 economic sectors represented in the ABM. The damage-scenario generator simulates a shock to individual agents in the ABM, which subsequently alter their behavior and create higher-order indirect effects over a given time period (see SI).

\section*{Acknowledgments} We acknowledge support from the EC H2020 project SmartResilience under grant agreement No 700621. Computations were performed in part on the Vienna Scientific Cluster. We would like to thank Prof. Georg Pflug as well as Dr. Anna Timonina-Farkas for helpful comments in designing the modeling approach and providing useful inputs.

\bibliography{econophysics,literature}

\end{document}